\documentclass[onecolumn,12pt]{IEEEtran}

\usepackage{amsmath}
\usepackage{amsfonts}
\usepackage{pifont}
\usepackage{amssymb}
\usepackage{amsthm}
\usepackage{tikz}
\usepackage{graphicx}
\usepackage{array}
\usepackage[small,justification=centering]{caption}

\theoremstyle{plain}

\newtheorem{lemma}{Lemma}
\newtheorem{proposition}{Proposition}
\newtheorem{definition}{Definition}
\newtheorem{theorem}{Theorem}
\newtheorem{corollary}{Corollary}
\theoremstyle{definition}
\newtheorem{example}{Example}

\begin{document}

\title{Repair Locality with Multiple Erasure Tolerance}

\author{\IEEEauthorblockN{Anyu~Wang and
        Zhifang~Zhang}

\IEEEauthorblockA{Key
Laboratory of Mathematics Mechanization, NCMIS\\
Academy of Mathematics and Systems Science, Chinese Academy of Sciences\\
Beijing, 100190\\
Email: wanganyu@amss.ac.cn,~ zfz@amss.ac.cn}
}
\maketitle
\thispagestyle{empty} 

\begin{abstract}
In distributed storage systems, erasure codes with locality $r$ is preferred because a coordinate can be recovered by accessing at most $r$ other coordinates which in turn greatly reduces the disk I/O complexity for small $r$.
However, the local repair may be ineffective when some of the $r$ coordinates accessed for recovery are also erased.

To overcome this problem, we propose the $(r,\delta)_c$-locality providing $\delta -1$ local repair options for  a coordinate.
Consequently, the repair locality $r$ can tolerate $\delta-1$ erasures in total.
We  derive an upper bound on the minimum distance $d$ for any linear $[n,k]$ code with information $(r,\delta)_c$-locality.
For general parameters, we prove existence of the codes that attain this bound when $n\geq k(r(\delta-1)+1)$, implying tightness of this bound.
Although the locality $(r,\delta)$ defined by Prakash et al provides the same level of locality and local repair tolerance as our definition,  codes with $(r,\delta)_c$-locality are proved to have more advantage in the minimum distance.
In particular, we construct a class of codes with all symbol $(r,\delta)_c$-locality where the gain in minimum distance is $\Omega(\sqrt{r})$ and the information rate is close to $1$.
\end{abstract}

\section{Introduction}

In distributed storage systems, using erasure codes instead of straightforward replication may lead to desirable improvements in storage overhead and reliability \cite{Compare}.
A challenge of the coding technique is to efficiently repair the packets loss caused by node failures so that the system keeps the same level of redundancy.
However, traditional erasure codes are inefficient in concern with the repair bandwidth as well as the number of disk accesses during the repair process. As an improvement, {\it regenerating codes} and codes with {\it repair locality} are proposed respectively. We focus on the latter in this paper.

As proposed by Gopalan et al \cite{Gopalan2012}, the $i$-th coordinate of an $[n,k,d]_q$ linear code has repair locality $r$, if the value at this coordinate of any codeword can be recovered by accessing at most  $r$ other coordinates. Applying to a distributed storage system in a way that each node stores a coordinate of the codeword, the code with repair locality $r \ll k$ is much more desirable because of its low disk I/O complexity for repair. Given $k,r$ and $d$, a lower bound on the codeword length is derived \cite{Gopalan2012}, and codes which are optimal with respect to this bound are also constructed \cite{Dimakis2012,Pyramid2007}.

However, for these locally repairable codes \cite{Dimakis2012,Pyramid2007,SimpleRC2012}, a problem rises when there are multiple node failures in the system.
Particularly, because only one local repair option for the  locality $r$ of a node (say $i$) is provided, if one of these $r$ nodes also fails, then node $i$ can no longer be repaired by accessing at most $r$ other nodes. That is, the repair locality $r$ can tolerate only one node failure.
Nevertheless, in today's large-scale distributed storage systems, multiple node failures are the norm rather than exception.
This motivates our pursuit of codes with multiple erasure tolerance for repair locality and also with other good properties.
The following example gives us some direction.

\begin{example} \label{ex_code7,3,4}
Consider a binary $[n=7,k=3,d=4]$ linear code with generator matrix
$$G = \begin{pmatrix} 1 & 0 & 0 & 0 & 1 & 1 & 1 \\ 0 & 1 & 0 & 1 & 0 & 1 & 1 \\ 0 & 0 & 1 & 1 & 1 & 0 & 1 \end{pmatrix}.$$
As displayed in Fig. \ref{FigOf[7,3,4]Code}, in the plane with seven points and seven lines (including the circle), each point is associated with a column vector of $G$, then the three vectors associated with collinear points add up to zero.
Thus the code has the following properties about repair locality.
\begin{itemize}
\item[(1)] Each coordinate has repair locality $r=2$.
\item[(2)] The repair locality $r$ of each coordinate can tolerate up to three erasures.
\end{itemize}
\end{example}
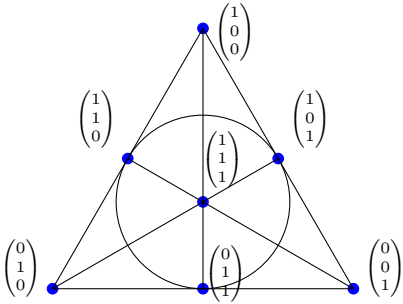
\begin{figure}[ht]
\centering
\begin{tikzpicture}[scale=2]
\filldraw [blue]
(0,0) circle (1pt)
(-1,0) circle (1pt)
(1,0) circle (1pt)
(-0.5,3^0.5/2) circle (1pt)
(0.5,3^0.5/2) circle (1pt)
(0,3^0.5) circle (1pt)
(0,3^0.5/3) circle (1pt);

\draw
(-1,0) -- (1,0)
(-1,0) -- (0,3^0.5)
(1,0) -- (0,3^0.5)
(-1,0) -- (0.5,3^0.5/2)
(1,0) -- (-0.5,3^0.5/2)
(0,0) -- (0,3^0.5);

\draw (0,3^0.5/3) circle (3^0.5/3);

\node [below = 2pt] at (0.15,0.4) {\tiny $\begin{pmatrix} 0\\1\\1\\ \end{pmatrix}$};
\node [below left] at (-1,0.4) {\tiny $\begin{pmatrix} 0\\1\\0\\ \end{pmatrix}$};
\node [below right] at (1,0.4) {\tiny $\begin{pmatrix} 0\\0\\1\\ \end{pmatrix}$};
\node [above right] at (0,3^0.5/1.2) {\tiny $\begin{pmatrix} 1\\0\\0\\ \end{pmatrix}$};
\node [above left] at (-0.5,3^0.5/2) {\tiny $\begin{pmatrix} 1\\1\\0\\ \end{pmatrix}$};
\node [above right] at (0.5,3^0.5/2) {\tiny $\begin{pmatrix} 1\\0\\1\\ \end{pmatrix}$};
\node [above right = 1pt] at (-0.1,3^0.5/3) {\tiny $\begin{pmatrix} 1\\1\\1\\ \end{pmatrix}$};
\end{tikzpicture}
\caption{The projective plane corresponding to the $[7,3,4]$ code.}
\label{FigOf[7,3,4]Code}
\end{figure}
 We compare the above code with some other codes which can also tolerate multiple erasures for local repair.

First, Prakash et al \cite{Prakash2012} define the locality $(r,\delta)$ by using a punctured subcode of length at most $r+\delta-1$. Since the subcode has minimum distance at least $\delta$, the repair locality $r$ can tolerate up to $\delta-1$ erasures. They derive a lower bound on the codeword length  under their definition of locality, i.e.
\begin{equation}\label{r,/delta bound 1}
n \geq d+ k - 1 + (\lceil \frac{k}{r} \rceil - 1)(\delta -1 )\;.
\end{equation}
Considering the code in Example \ref{ex_code7,3,4}, we set $k=3,d=4,r=2,\delta=4$, and get
$n \ge 4+3 -1 +(\lceil \frac{3}{2} \rceil -1)(4-1) = 9$ from the bound (\ref{r,/delta bound 1}).
But actually the code in Example \ref{ex_code7,3,4} has length $7$, which outperforms the bound (\ref{r,/delta bound 1}).

Another comparison is with the minimum-bandwidth regenerating code based on an inner fractional repetition code which can perform exact uncoded repair even under multiple node failures. In \cite{FractionalRepetitionCodes2010} they build such a code of length $7$ based on the same projective plane as in Fig. \ref{FigOf[7,3,4]Code}. Suppose the original data is of size $B$, Table \ref{t1} displays some comparisons between these two codes. We can see the code in Example \ref{ex_code7,3,4} outperforms the code of \cite{FractionalRepetitionCodes2010} in both storage overhead and repair locality. Moreover, the repair locality of the former code can tolerate one erasure more than that of the latter.
\begin{table}[ht]
\centering
\caption{}\label{t1}
\renewcommand{\arraystretch}{1.3}
\begin{tabular}{ | p{2.5cm}<{\centering} |  p{2.3cm}<{\centering} | p{2.3cm}<{\centering} |}
\hline
 & code in Example \ref{ex_code7,3,4} & code in \cite{FractionalRepetitionCodes2010} \\ \hline
storage per node & $\frac{1}{3} B$ & $\frac{1}{2} B$  \\ \hline
repair locality & 2 & 3 \\ \hline
local repair tolerance  & 3  & 2 \\ \hline
repair bandwidth & $\frac{2}{3} B$  & $\frac{1}{2} B$  \\ \hline
\end{tabular}
\end{table}

In summary, the code in Example \ref{ex_code7,3,4} has many appealing properties: binary code, low repair locality, high local repair tolerance, and shorter codeword length (or larger minimum distance). This encourages us to study a new kind of repair locality similar to that of this code.

\subsection{Our Results}
For any $[n,k,d]_q$ linear code, we define the repair locality from a combinatorial perspective, denoted as $(r,\delta)_c$-locality. The main idea is to guarantee $\delta-1$ \footnote{We set the tolerance as $\delta-1$ in order to make it consistent with that of the locality $(r,\delta)$ defined in \cite{Prakash2012}. } repair options for the locality $r$, therefore the failed node can still be locally repaired by accessing at most $r$ other nodes as long as the total number of erasures is no more than $\delta-1$. For a linear code whose information symbols have the $(r,\delta)_c$-locality, we prove a lower bound on the codeword length (or equivalently, an upper bound on the minimum distance),
\begin{equation*}
n \geq d+k -1 + \mu,
\end{equation*}
where $\mu = \lceil \frac{(k-1)(\delta -1) +1}{(r-1)(\delta -1) +1} \rceil -1$.
It can be verified that the code in Example \ref{ex_code7,3,4} satisfies the $(r,\delta)_c$-locality with $r=2, \delta=4$, and meets the bound with equality.
We further prove the existence of codes with information $(r,\delta)_c$-locality that attain the above bound for any $r,\delta,k$ and the length $n\ge k(r(\delta-1)+1)$, which indicates tightness of the bound in general.

Comparing with the bound (\ref{r,/delta bound 1}), we show by detailed computation that under the same locality $r$ and local repair tolerance $\delta -1$, an $[n,k]$ code with $(r,\delta)_c$-locality outperforms the one with locality $(r,\delta)$ in terms of the minimum distance. This advantage is further certified through some specific codes presented later. In particular, we build a class of codes with all symbol $(r,\delta)_c$-locality where the gain in minimum distance is $\Omega(\sqrt{r})$ and the information rate is close to 1.

\subsection{Related Works}
Some existing erasure codes for distributed storage also consider tolerating multiple erasures for local repair.

As started earlier, the locality $(r,\delta)$ defined in \cite{Prakash2012} takes advantage of inner-error-correcting codes, while our $(r,\delta)_c$-locality is defined in a combinatorial way.
This difference brings improvements in the codeword length and the minimum distance.
Detailed comparisons can be found in Section \ref{sec_tightness_and_compare} and \ref{sec_constrct}.

Paper \cite{FractionalRepetitionCodes2010} designed the minimum-bandwidth regenerating code based on an inner fractional repetition code.
It cares primarily about achieving minimum bandwidth and uncoded repair, rather than repair locality.

The metric ``local repair tolerance" was introduced in \cite{CodeMultipleRepairAlternatives2013} to measure the maximum number of erasures that do not compromise local repair.
A class of codes with high local repair tolerance and low repair locality,  named $pg$-BLRC code, was designed there.
It further gave the information rate region of such codes.
However, the construction of high rate $pg$-BLRC codes depends on a special class of partial geometry named generalized quadrangle, of which only a few instances are known until now.
Our $(r,\delta)_c$-locality is similar to the metric ``local repair tolerance", while we seek more general constructions of codes that have good properties in repair locality, information rate and fault-tolerance.
\subsection{Organization}
In Section \ref{sec_bound_tightness} we formally define the $(r,\delta)_c$-locality and prove a lower bound on the codeword length.
Then Section III states this lower bound can be attained by some codes with general parameters.
A comparison with locality $(r,\delta)$ is given by detailed computation.
Section \ref{sec_constrct} provides some constructions of codes with $(r,\delta)_c$-locality and Section V concludes the paper.

\section{Definition and Lower Bound}\label{sec_bound_tightness}

Let $\mathcal{C}$ be an $[n,k,d]_q$ linear code with generator matrix
$ G = ( g_1 , \cdots , g_n)$,
where $g_i \in \mathbb{F}_q^k$ is a column vector for $i=1,\cdots,n$.
Then a message $x \in \mathbb{F}_q^k$ is encoded into
$$ x^\tau G = (x^\tau g_1, \cdots,x^\tau g_n) .$$
Denote $[t]=\{1,2,\cdots,t\}$ for any positive integer $t$.
Given $\mathcal{C}$ and the matrix $G$, we introduce the following notations and concepts:
\begin{itemize}
\item[(1)] For any set $N \subseteq [n]$, let $\text{span} (N) $ be the linear space spanned by $\{g_i | i \in N\}$ over $\mathbb{F}_q$.
\item[(2)] For any set $N \subseteq [n]$, let $\text{rank} (N)$ be the dimension of $\text{span} (N) $.
\item[(3)] A set $I \subseteq [n]$ is called an information set if $\left| I \right| = \text{rank}(I)=k$.
\end{itemize}

The following lemma describes a useful fact about the $[n,k,d]_q$ linear code $\mathcal{C}$.
Its proof comes from basic concepts of linear codes \cite{CodingTheory}.

\begin{lemma}\label{rank(n-d+1)=k}
For an $[n,k,d]_q$ linear code $\mathcal{C}$, let $N \subseteq [n]$ have the maximum size among all the subsets with rank less than $k$,  then $d = n - |N|$.
\end{lemma}

\begin{definition}\label{def_of_r,delta_codes}
For $1 \le i \le n$, the $i$-th coordinate of an $[n,k,d]_q$ linear code $\mathcal{C}$ is said to have $(r,\delta)_c$-locality if there exists $\delta-1$ pairwise disjoint sets $R^{(i)}_1,\cdots,R^{(i)}_{\delta -1} \subseteq [n] \backslash \{i\}$, called repair sets, satisfying for $1\leq \xi \leq \delta-1$,
\begin{itemize}
\item[(1)] $|R^{(i)}_\xi| \le r$, and

\item[(2)] $g_i \in \text{span} (R^{(i)}_\xi)$.
\end{itemize}
\end{definition}

It is clear that the $(r,\delta)_c$-locality ensures repair locality $r$ and the tolerance of $\delta-1$ erasures for this repair locality.
We call the code $\mathcal{C}$ has information $(r,\delta)_c$-locality if there is an information set $I$ such that for any $i \in I$, the $i$-th coordinate has $(r,\delta)_c$-locality. Similarly,
$\mathcal{C}$ has all symbol $(r,\delta)_c$-locality if for any $i \in [n]$, its $i$-th coordinate has $(r,\delta)_c$-locality.
Note that $r =1$ implies repetition and $\delta = 1$ means no locality, therefore we only consider codes with $r,\delta \ge 2$.
Additionally, we always assume $r < k$ because an MDS code is optimal for the locality $r \ge k$.

Given $k,d,r$ and $\delta$, our goal is to minimize the codeword length $n$.
The following theorem provides a lower bound of $n$  for codes with information $(r,\delta)_c$-locality.

\begin{theorem}\label{boundthm}
For any $[n,k,d]_q$ linear code with information $(r,\delta)_c$-locality,
\begin{eqnarray}\label{generalbound}
n \ge d+k -1+ \mu ,
\end{eqnarray}
where $\mu = \lceil \frac{(k-1)(\delta-1)+1}{(r-1)(\delta-1)+1} \rceil -1 $.
\end{theorem}

\begin{IEEEproof}
It is equivalent to prove
$d \le n- (k-1+\mu)$.
From Lemma \ref{rank(n-d+1)=k}, we prove it by constructing a set $S_l\subseteq [n]$ such that $ |S_l|\ge k-1 + \mu$ and $\text{rank}(S_l)<k$.

Let $I$ be an information set such that each coordinate in $I$ has $(r,\delta)_c$-locality.
For any $i \in I$ and $0\le\xi\le\delta-1$, denote $N^{(i)}_\xi = \{i\}\cup R_1^{(i)}\cup \cdots \cup R_\xi^{(i)}$, then
$$\text{rank} (N_\xi^{(i)}) \le (r-1)\xi+1,$$
since $g_i\in\bigcap_{j=1}^\xi \text{span}(R_j^{(i)})$ and increase of the rank caused by adding $R_j^{(i)}$ is less than $r-1$ for $1\leq j\leq \xi$.

The set $S_l$ is constructed by the following algorithm:

\vspace{8pt}
\begin{itemize}
\item[1.]Set $h=1$ and $S_0 = \{\}$.
\item[2.]While $\text{rank} (S_{h-1}) \le k-2$ :
\item[3.] \hspace*{9.5pt}Pick $i \in I $ such that $g_i \notin \text{span}(S_{h-1})$.
\item[4.] \hspace*{9.5pt}If $\text{rank} (S_{h-1} \cup N^{(i)}_{\delta-1}) < k$, set $S_h = S_{h-1} \cup N^{(i)}_{\delta-1}$.
\item[5.] \hspace*{9.5pt}Else pick $\theta \in [0,\delta-1)$ and $R \subseteq R_{\theta +1}^{(i)}$ such that
\item[] \hspace*{16pt}$\text{rank} (S_{h-1} \cup N^{(i)}_{\theta}) < k$,
\item[] \hspace*{16pt}$\text{rank} (S_{h-1} \cup N^{(i)}_{\theta +1}\} \ge  k$
\item[] \hspace*{16pt}and $\text{rank} (S_{h-1} \cup N^{(i)}_{\theta} \cup R ) = k-1$.
\item[6.] \hspace*{16pt}Set $S_h = S_{h-1}\cup N^{(i)}_{\theta} \cup R$.
\item[7.] \hspace*{9.5pt}$h=h+1$.
\end{itemize}
\vspace{8pt}

Note that in step 3 the desired $i$ exists since $\text{rank} (S_{h-1}) \le k-2$ and $\text{rank}(I) = k$.
Let $S_l$ be the set with which the algorithm terminates.
Then we can see that $\text{rank} (S_l) = k-1$.
Next, we estimate the size of $S_l$.
For $1\leq h\leq l$, define
$$s_h = | S_h| - |S_{h-1}|$$
and
$$t_h = \text{rank} (S_h) - \text{rank} (S_{h-1}),$$
i.e. the increase of $S_h$ in the size and rank respectively.
Then
$$ |S_l| = \sum_{h=1}^{l} s_h,\text{ } \text{rank} (S_l) = \sum_{h=1}^{l} t_h = k-1.$$
Since $S_l$ may be generated at Step 4 or Step 6 of the algorithm, we consider the two cases respectively.

Case 1. $S_l$ is generated at Step 4.
Then we have
$$ l \ge \lceil \frac{k-1}{(r-1)(\delta-1)+1} \rceil,$$
because $k-1 = \sum_{h=1}^{l} t_h$ and
\begin{eqnarray*}
t_h &=& \text{rank} (S_h) - \text{rank} (S_{h-1}) \\
& \le& \text{rank} (N^{(i)}_{\delta-1}) \\
& \le& (r-1)(\delta-1)+1.
\end{eqnarray*}
For any $i\in I$, since vector $g_{i}$ lies in the intersection of the $\delta -1$ spaces $\text{span} (R_1^{(i)}),\cdots,\text{ span} (R_{\delta-1}^{(i)})$, adding $N^{(i)}_{\delta-1}$ to $S_{h-1}$ makes increase of the rank less than increase of the set size by at least $\delta-1$, namely,
$$t_h\le s_h-(\delta-1)\;.$$
Thus
\begin{eqnarray*}
|S_l|&=& \sum_{h=1}^{l} s_h \\
&\ge& \sum_{h=1}^{l} t_h + l (\delta-1) \\
&\ge& k-1 + \lceil \frac{k-1}{(r-1)(\delta-1)+1} \rceil (\delta-1) \\
&\ge& k-1 + \mu,
\end{eqnarray*}
where the last inequality holds because
\begin{eqnarray}\label{eq1_in_proof}
\mu & = & \lceil \frac{(k-1)(\delta-1)+1}{(r-1)(\delta-1)+1}\rceil -1 \nonumber \\
& = & \lfloor \frac{(k-1)(\delta-1)}{(r-1)(\delta-1)+1} \rfloor \nonumber \\
& \le & \lceil \frac{k-1}{(r-1)(\delta-1)+1} \rceil (\delta-1).
\end{eqnarray}

Case 2. Suppose $S_l$ is generated at Step 6.
Then it has $\text{rank} (S_{l-1}\cup N^{(i)}_{\delta-1}) =k$.
Similarly we have $l \ge \lceil \frac{k}{(r-1)(\delta-1)+1} \rceil$.
For $1\leq h\leq l-1$, it also has
$$t_h \le s_h -(\delta-1).$$
Particularly,
$$t_{l} \le s_{l} - \theta.$$
Thus
\begin{eqnarray} \label{eq2_in_proof}
|S_{l}| &=&  \sum_{h=1}^{l} s_h \nonumber \\
&\ge& \sum_{h=1}^{l} t_h + (l -1) (\delta-1) + \theta \nonumber \\
&= & k-1 + (l -1) (\delta-1) + \theta
\end{eqnarray}

If $l \ge \lceil \frac{k}{(r-1)(\delta-1)+1} \rceil +1 $,
\begin{eqnarray*}
|S_l| & \ge & k-1+ \lceil \frac{k}{(r-1)(\delta-1)+1} \rceil (\delta-1) + \theta \\
& \ge & k-1+\mu,
\end{eqnarray*}
where the last inequality follows from $\theta \ge 0$ and (\ref{eq1_in_proof}).

If $l = \lceil \frac{k}{(r-1)(\delta-1)+1} \rceil$, note that $\theta$ is chosen such that
$$\text{rank} (S_{l-1} \cup N^{(i)}_{\theta +1}) \ge k.$$
On the other hand,
$$\text{rank} (S_{l -1}) \le (l -1)((r-1)(\delta-1)+1).$$
Then
\begin{eqnarray*}
k &\le& \text{rank} (S_{l-1} \cup N^{(i)}_{\theta +1}) \\
&\le& (l -1)((r-1)(\delta-1)+1) + \text{rank} (N^{(i)}_{\theta +1}) \\
&\le& (l -1)((r-1)(\delta-1)+1) + (\theta+1)(r-1)+1.
\end{eqnarray*}
It follows that
$$ \theta \ge (\lceil \frac{1}{r-1} (k-1 - (l-1)((r-1)(\delta-1)+1)) \rceil-1)^+,$$
where $t^+ = \max\{t,0\}$ for any integer $t$.

Let $k = \alpha ((r-1)(\delta -1)+1) +\beta$, where $\alpha$ and $\beta$ are integers and $1\le \beta \le (r-1)(\delta-1)+1$, then $l = \alpha +1$ and $\theta \ge (\lceil \frac{\beta-1}{r-1}\rceil-1)^+$.
Thus (\ref{eq2_in_proof}) implies that
\begin{eqnarray*}
\left| S_l \right| & \ge & k-1 + (l-1)(\delta-1)+\theta \\
& \ge & k -1 + \alpha (\delta -1) + (\lceil \frac{\beta-1}{r-1}\rceil-1)^+ \\
& \ge & k-1 + \mu,
\end{eqnarray*}
where the last inequality follows from
\begin{eqnarray*}
\mu & = & \lceil \frac{(k-1)(\delta-1)+1}{(r-1)(\delta-1)+1} \rceil -1 \\
& = & \alpha (\delta-1) + \lceil \frac{(\beta-1)(\delta-1)+1}{(r-1)(\delta-1)+1} \rceil -1\\
& \le & \alpha(\delta -1) + (\lceil \frac{\beta-1}{r-1}\rceil-1)^+.
\end{eqnarray*}
\end{IEEEproof}

We say a linear code with information $(r,\delta)_c$-locality is optimal if the bound (\ref{generalbound}) is satisfied with equality.
The code in Example \ref{ex_code7,3,4} is optimal in this sense.
We will give more optimal codes in the rest of this paper.

\section{Tightness of the bound}\label{sec_tightness_and_compare}
In this section, we certify tightness of the bound (\ref{generalbound}) by giving existence of a class of optimal codes with general parameters.
Then we compare bound (\ref{generalbound}) with bound (\ref{r,/delta bound 1})  showing the advantage of $(r,\delta)_c$-locality over the locality $(r,\delta)$ of \cite{Prakash2012} in the minimum distance.
\begin{theorem}\label{thm_tightness}
If $q \ge 1 + \binom{n}{k+\mu}$ and $n = k(r(\delta-1)+1)$, then there exists an optimal $[n,k,d]_q$ linear code with information $(r,\delta)_c$-locality.
\end{theorem}
\begin{IEEEproof}
For $1 \le l \le  k$ and $ 1\le a \le \delta -1$, let $B^{(l)}_a = \{ s^{(l)},s^{(l)}_{a1},\cdots,s^{(l)}_{ar}\}$ be a set of $r+1$ points.
Denote $N_l = B^{(l)}_1 \cup \cdots \cup B^{(l)}_{\delta - 1}$ and $N=\cup_{l=1}^kN_l$.
Thus $N$ is a set of $n$ points and Fig. \ref{fig_N} gives a graphical elaboration of the points.
Specifically, each point in the graph denotes a coordinate of the code and thus a column of the generator matrix.
The $r+1$ points of $B_a^{(l)}$ lie in a line in the graph meaning linear dependence among these coordinates.
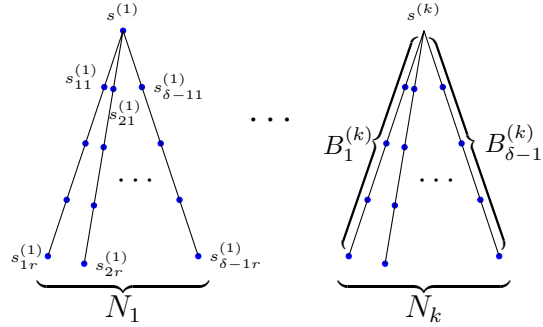
\begin{figure}[htbp]
\centering
\begin{tikzpicture}

\filldraw[blue]
(0,0) circle (1pt);

\foreach \x in {1,1/4,2/4,3/4}
{\filldraw[blue] (-\x,-3 * \x) circle (1pt);
\filldraw[blue] (\x,-3 * \x) circle (1pt);
\filldraw[blue] (-3.1/6 * \x,-3.1 * \x) circle (1pt);}

\node [above left] at (-0.7/4,-3.7/4) { \tiny $s^{(1)}_{11}$};
\node [left] at (-0.9,-3) { \tiny $s^{(1)}_{1r}$};
\node [below] at (-3.1/24 + 0.16,-3.1/4) {\tiny $s^{(1)}_{21}$};
\node [right] at (-3.1/6,-3.1) {\tiny $s^{(1)}_{2r}$};
\node [right] at (1/4,-3/4) {\tiny $s^{(1)}_{\delta-1 1}$};
\node [right] at (1,-3) {\tiny $s^{(1)}_{\delta -1 r}$};

\draw
(0,0) -- (-1,-3)
(0,0) -- (1,-3)
(0,0) -- (-3.1/6,-3.1);

\node at (.2,-2) {$\cdots$};
\node [above] at (0,0) {\tiny $s^{(1)}$};
\node at (0,-3.4) {\scriptsize $\underbrace{\hspace*{65pt}}$};
\node at (0,-3.7) {$N_1$};

\node at (2,-1.2) {\large $\cdots$};

\foreach \x in {1,1/4,2/4,3/4}
{\filldraw[blue] (4 -\x,-3 * \x) circle (1pt);
\filldraw[blue] (4 + \x,-3 * \x) circle (1pt);
\filldraw[blue] (4 -3.1/6 * \x,-3.1 * \x) circle (1pt);}

\draw
(4,0) -- (3,-3)
(4,0) -- (5,-3)
(4,0) -- (4+ -3.1/6,-3.1);

\node at (4+ .2,-2) {$\cdots$};
\node [above] at (4+ 0,0) {\tiny $s^{(k)}$};
\node at (4+ 0,-3.4) {\scriptsize $\underbrace{\hspace*{65pt}}$};
\node [rotate = 251] at (3.38,-1.48) {\tiny $\underbrace{\hspace*{85pt}}$};
\node at (3,-1.5) {\small $B^{(k)}_1$};
\node [rotate = 109] at (4.61,-1.48) {\tiny $\underbrace{\hspace*{85pt}}$};
\node at (5.2,-1.5) {\small $B^{(k)}_{\delta-1}$};
\node at (4+ 0,-3.7) {$N_k$};

\end{tikzpicture}
\caption{The set $N$ of $n$ points.}
\label{fig_N}
\end{figure}

We claim that for $q \ge 1 + \binom{ n}{  k+ \mu}$ there exists a $k \times n$ matrix $G = (g_i)_{i \in N}$ over $\mathbb{F}_q$ satisfying the following three conditions.
\begin{itemize}
    \item[(1)] $\sum_{i \in B_a^{(l)}} g_i = 0$ for $1\leq a\leq \delta-1$ and $1\leq l\leq k$.
    \item[(2)] $\text{rank}(g_{s^{(1)}},\cdots,g_{s^{(k)}}) = k$.
    \item[(3)] For any $M \subseteq N$ with $\left| M \right| = k+\mu$, $\text{rank}(M) = k$.
\end{itemize}

The claim is proved in Proposition \ref{thm_tightness_restate} in the Appendix.
In fact, let $\mathcal{C}$ be a code with the generator matrix $G$, then the condition (1) and (2) guarantee that $I = \{ s^{(1)},\cdots,s^{(k)}\}$ is an information set and each information symbol has $(r,\delta)_c$-locality.
The condition (3) implies the minimum distance of $\mathcal{C}$ is at least $n-k+1-\mu$, which is deduced from Lemma \ref{rank(n-d+1)=k}.
Then by Theorem \ref{boundthm} the bound (\ref{generalbound}) is met with equality.
Hence we have constructed the optimal code with length $n=k(r(\delta-1)+1)$.
\end{IEEEproof}

Actually, we can add more independent columns to the above matrix $G$  as parities.
Then the condition (1) and (2) still hold.
And by further increasing the field size, the condition (3) also hold for  the matrix G with additional columns, which implies attainment of bound (2) for a code with larger length.
Therefore we can extend the construction to $n\geq k(r(\delta-1)+1)$ and get the following corollary.

\begin{corollary}
For $n\geq k(r(\delta-1)+1)$ and sufficiently large $q$, there exists an optimal $[n,k,d]_q$ linear code with information $(r,\delta)_c$-locality.
\end{corollary}

We next compare the two kinds of $(r,\delta)$ locality in terms of the minimum distance.
Equivalently, Theorem \ref{boundthm} gives an upper bound on the minimum distance, i.e.,
\begin{equation*}
d \le n-k+1 -\mu.
\end{equation*}
On the other hand, for codes with locality $(r,\delta)$ introduced in \cite{Prakash2012}, the minimum distance is upper bounded by
$$d \le n - k +1 - (\lceil \frac{k}{r} \rceil - 1)(\delta -1).$$
Then we have
\begin{eqnarray*}
\mu & = & \lceil \frac{(k-1)(\delta-1)+1}{(r-1)(\delta-1)+1} \rceil -1 \\
& = & \lceil \frac{(k-r)(\delta -1)}{(r-1)(\delta-1)+1} \rceil \\
& \le & \lceil \frac{k-r}{(r-1)(\delta-1)+1} \rceil (\delta -1) \\
& \le & \lceil \frac{k-r}{r} \rceil (\delta -1) \\
& = & (\lceil \frac{k}{r} \rceil - 1)(\delta -1).
\end{eqnarray*}

That is, optimal codes with $(r,\delta)_c$-locality always possess preferable minimum distance than codes with locality $(r,\delta)$.
Actually, in Section \ref{sec_constrct} we will give a class of codes with all symbol $(r,\delta)_c$-locality which have information rate close to $1$ and minimum distance exceeding codes with locality $(r,\delta)$ by $\Omega (\sqrt{r})$.

\section{Construction of codes with $(r,\delta)_c$-locality}\label{sec_constrct}
In this section, we present some constructions of codes with all symbol $(r,\delta)_c$-locality.
It is evident that the bound (\ref{generalbound}) proved for information locality also holds for all symbol locality.
\begin{example}\label{ex_6,3,3}
Consider the binary $[6,3,3]$ code with generator matrix
$$ G = \begin{pmatrix} 1 & 0 & 0 & 1 & 1 & 1 \\ 0 & 1 & 0 & 1 & 0 & 1 \\ 0 & 0 & 1 & 0 & 1 & 1 \end{pmatrix}. $$
Similar to Example \ref{ex_code7,3,4}, the code associates with the plane in Fig. \ref{fig_6,3,3} which is obtained by deleting a point and three lines from the plane in Fig. \ref{FigOf[7,3,4]Code}.  Consequently,  the code has information rate $\frac{1}{2}$ and all symbol $(r, \delta)_c$-locality with $r=2$ and $\delta=3$.

\begin{figure}[ht]
\centering
\begin{tikzpicture}[scale=1.7]

\filldraw [blue]
(-1,0) circle (1pt)
(1,0) circle (1pt)
(-0.5,3^0.5/2) circle (1pt)
(0.5,3^0.5/2) circle (1pt)
(0,3^0.5) circle (1pt)
(0,3^0.5/3) circle (1pt);

\draw
(-1,0) -- (0,3^0.5)
(1,0) -- (0,3^0.5)
(-1,0) -- (0.5,3^0.5/2)
(1,0) -- (-0.5,3^0.5/2);

\node [below left] at (-1,0.4) {\tiny $\begin{pmatrix} 0\\1\\0\\ \end{pmatrix}$};
\node [below right] at (1,0.4) {\tiny $\begin{pmatrix} 0\\0\\1\\ \end{pmatrix}$};
\node [above right] at (0,3^0.5/1.2) {\tiny $\begin{pmatrix} 1\\0\\0\\ \end{pmatrix}$};
\node [above left] at (-0.5,3^0.5/2) {\tiny $\begin{pmatrix} 1\\1\\0\\ \end{pmatrix}$};
\node [above right] at (0.5,3^0.5/2) {\tiny $\begin{pmatrix} 1\\0\\1\\ \end{pmatrix}$};
\node [above] at (0,3^0.5/3) {\tiny $\begin{pmatrix} 1\\1\\1\\ \end{pmatrix}$};

\end{tikzpicture}
\caption{The graph corresponding to the $[6,3,3]$ binary code}
\label{fig_6,3,3}
\end{figure}
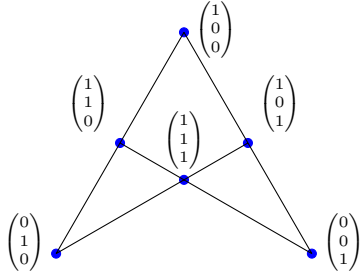

We then show the code is optimal with respect to bound (\ref{generalbound}).
Since $ \mu = \lceil \frac{(k-1)(\delta -1)+1}{(r-1)(\delta-1)+1}\rceil -1 =1$ in this case, bound (\ref{generalbound}) indicates
$$n \ge d+k -1 +\mu  =6.$$
Therefore, the bound is met with equality.

However, fix $k=3$ and $d=3$, under the same level of locality and local repair tolerance, bound (\ref{r,/delta bound 1}) indicates that a code with locality $(r=2,\delta=3)$ has length $n \ge d+k-1+(\lceil\frac{k}{r}\rceil-1)(\delta-1)=7$.

Though the codes in Example \ref{ex_code7,3,4} and \ref{ex_6,3,3} are both optimal with respect to bound (\ref{generalbound}), their information rate are no more than $\frac{1}{2}$.
In the following we will give a class of codes which have information rate close to $1$ and are near optimal with respect to bound (\ref{generalbound}).
\end{example}

\begin{example}\label{ex_square_codes}
Let $r$ be a positive integer, $n=(r+1)^2$, and $r+1\leq k\leq r^2$.
We next construct $[n,k]$ linear codes with all symbol $(r,\delta)_c$-locality.

Let $X = \{ x_{i,j} \}_{1 \le i ,j \le r+1}\subseteq\mathbb{F}_q^k$ be a set of $(r+1)^2$ column vectors such that
\begin{equation}\label{eq55}
\begin{cases}
\sum_{i=1}^{r+1} x_{i,j} = 0, \text{ for } 1\le j \le r+1 \\
\sum_{j=1}^{r+1} x_{i,j} = 0, \text{ for } 1 \le i \le r+1.
\end{cases}
\end{equation}
In fact, the $(r+1)^2$ vectors can be chosen as follows.
First, we randomly choose $r^2$ vectors $\{x_{i,j}\}_{1\le i,j\le r}$.
Then let $x_{i,r+1} = - \sum_{j=1}^r x_{i,j}$ for $1 \le i \le r$, and $x_{r+1, j} = - \sum_{i=1}^r x_{i,j}$ for $1 \le j \le r+1$. It can be verified the condition (\ref{eq55}) is satisfied.

There is a grid corresponding to $X$.
As in Fig. \nolinebreak \ref{fig_square_code}, vector $x_{i,j}$ stands for the cross point of the $i$-th row and the $j$-th column in the grid.
The sum of all $r+1$ vectors in the same row (or the same column) is zero.

\begin{figure}[htbp]
\centering
\begin{tikzpicture}[scale=0.68]
\filldraw [blue]
(0,0) circle (2pt)
(0,1) circle (2pt)
(1,0) circle (2pt)
(1,1) circle (2pt)
(0,4) circle (2pt)
(0,5) circle (2pt)
(1,4) circle (2pt)
(1,5) circle (2pt)
(4,0) circle (2pt)
(5,0) circle (2pt)
(4,1) circle (2pt)
(5,1) circle (2pt)
(4,4) circle (2pt)
(5,4) circle (2pt)
(4,5) circle (2pt)
(5,5) circle (2pt);

\draw
(0,0) -- (0,1.7)
(1,0) -- (1,1.7)
(4,0) -- (4,1.7)
(5,0) -- (5,1.7)
(0,3.3) -- (0,5)
(1,3.3) -- (1,5)
(4,3.3) -- (4,5)
(5,3.3) -- (5,5)
(0,0) -- (1.7,0)
(0,1) -- (1.7,1)
(0,4) -- (1.7,4)
(0,5) -- (1.7,5)
(3.3,0) -- (5,0)
(3.3,1) -- (5,1)
(3.3,4) -- (5,4)
(3.3,5) -- (5,5);
\draw [dashed]
(0,1.7) -- (0,3.3)
(1,1.7) -- (1,3.3)
(4,1.7) -- (4,3.3)
(5,1.7) -- (5,3.3)
(1.7,0) -- (3.3,0)
(1.7,1) -- (3.3,1)
(1.7,4) -- (3.3,4)
(1.7,5) -- (3.3,5);

\node [below left] at (0,0) {\small $x_{r+1,1}$};
\node [below] at (1,0) {\small $x_{r+1,2}$};
\node [below] at (4,0) {\small $x_{r+1,r}$};
\node [below right] at (5,0) {\small $x_{r+1,r+1}$};

\node [left] at (0,1) {\small $x_{r,1}$};
\node [above right] at (1,1) {\small $x_{r,2}$};
\node [above right] at (4,1) {\small $x_{r,r}$};
\node [right] at (5,1) {\small $x_{r,r+1}$};

\node [left] at (0,4) {\small $x_{2,1}$};
\node [above right] at (1,4) {\small $x_{2,2}$};
\node [above right] at (4,4) {\small $x_{2,r}$};
\node [right] at (5,4) {\small $x_{2,r+1}$};

\node [above left] at (0,5) {\small $x_{1,1}$};
\node [above] at (1,5) {\small $x_{1,2}$};
\node [above] at (4,5) {\small $x_{1,r}$};
\node [above right] at (5,5) {\small $x_{1,r+1}$};

\end{tikzpicture}
\caption{The grid corresponding to vectors $\{x_{i,j}\}$.}
\label{fig_square_code}
\end{figure}
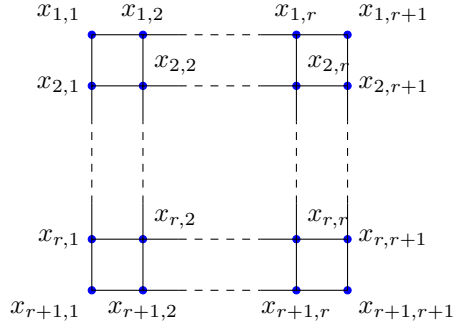

Consider an $[n,k]$ code $\mathcal{C}$ with the generator matrix $G(X)$ consisting of the $(r+1)^2$ vectors in $X$ as column vectors.
Then $\mathcal{C}$ clearly has all symbol $(r,\delta = 3)_c$-locality.
Specifically, each cross point in the grid stands for a coordinate of  $\mathcal{C}$.
Thus each coordinate lies in a row  (and a column) of the grid together with other $r$ coordinates which constitute the local repair sets for that coordinate.
We call $\mathcal{C}$ a square code with locality $r$.

In the following, we estimate the minimum distance $d$ of $\mathcal{C}$.
Firstly, for $x\in [2r+1]$, define
\begin{equation*}
f(x) = \begin{cases}
x(r+1) - \frac{x^2}{4}, \text{ if } 2 \mid x \\
x(r+1) - \frac{x^2 -1}{4}, \text{ if } 2 \nmid x,
\end{cases}
\end{equation*}
then let
$$\mu_k = \max \{ x | f(x) -x \le k-1\}.$$
Note that $\mu_k$ is well defined because $f(x) -x $ is an increasing function with respect to $x$ and
\begin{equation*}
\begin{cases} f(0)-0 = 0 < k-1 \\ f(2r+1) - (2r+1) = r^2 \ge k. \end{cases}
\end{equation*}

We then prove that for $q > \binom{n}{k+\mu_k}$ there exists a generator matrix $G(X)$ over $\mathbb{F}_q$ such that the minimum distance of $\mathcal{C}$ satisfies
$$d \ge n-k+1-\mu_k.$$
See Proposition \ref{prop_distance_ex3} in the Appendix for proof details.

When $r+1 \le k \le 2r-1$, it can be deduced $\mu = \mu_k =1$.
Therefore the square code is optimal with respect to bound (\ref{generalbound}) in this case.

\begin{figure}[tbp]
\centering

\includegraphics[width = 0.68 \textwidth]{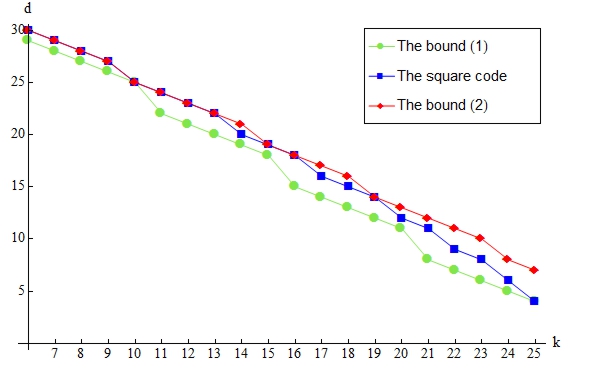}

\caption{The comparing of the three codes for $r = 5$.}
\label{Fig_comparing_three_codes}
\end{figure}

In other cases, the square code may not attain bound (\ref{generalbound}), but can always outperform the bound (\ref{r,/delta bound 1}).
For example, Fig. \nolinebreak \ref{Fig_comparing_three_codes} displays three curves indicating the $(k,d)$ pairs' respectively for the square code, the bound (\ref{r,/delta bound 1}) and the bound (\ref{generalbound}) at the parameters  $r=5$ and $n=36$.

Particularly, the minimum distance gap between the square code and the bound (\ref{r,/delta bound 1}) can be $\Omega (\sqrt{r})$.
For example, let $n=(r+1)^2, k = r^2-r+1$ and $\delta =3$, then the minimum distance of the square code satisfies
$$d \ge n-k+1 -\mu_k,$$
where $\mu_k \le 2(r-\lfloor \sqrt{r-1} \rfloor) - 1$ because
\begin{eqnarray*}
& & f(2(r-\lfloor \sqrt{r-1} \rfloor)) - 2(r-\lfloor \sqrt{r-1} \rfloor) \\
& = & r^2 - (\lfloor \sqrt{r-1} \rfloor)^2 \\
& \ge & r^2 -r+1\\
& = & k.
\end{eqnarray*}
On the other hand, the bound (\ref{r,/delta bound 1}) indicates that
\begin{eqnarray*}
d & \le & n-k+1 - (\lceil \frac{k}{r} \rceil -1)( \delta -1)\\
& = & n-k+1 - 2(r-1).
\end{eqnarray*}
Therefore the gap is no less than
\begin{eqnarray*}
& &( n-k+1 - \mu_k) - (n-k+1 - (\lceil\frac{k}{r}\rceil-1)(\delta-1))  \\
& \ge & 2(\lfloor \sqrt{r-1} \rfloor -1) +1 \\
& = & \Omega (\sqrt{r}) .
\end{eqnarray*}
Meanwhile, we note for $k=r^2-r+1$ the square code has information rate approaching $1$ as $r$ grows.

\end{example}

\section{Conclusions}
The $(r,\delta)_c$-locality proposed in this paper guarantees $\delta-1$ erasure tolerance for local repair in a combinatorial way. It brings improvement in the minimum distance comparing with the locality $(r,\delta)$ which provides multiple erasure tolerance for locality by using inner-error-correcting codes. We derive a lower bound on the codeword length for codes with information $(r,\delta)_c$-locality and prove the existence of codes attaining this bound with general parameters. Moreover, we present some specific codes with all symbol $(r,\delta)_c$-locality which are optimal with respect to the bound. In particular, the square code in Example \ref{ex_square_codes} has information rate approaching $1$ and is near optimal with respect to the bound. Actually, considering the specific structure of the repair sets, we can get a refined bound of the codeword length for codes with $(r,\delta)_c$-locality, and the square code can be proved attaining this refined bound. We leave the details in another paper.

\section*{Appendix}
\begin{proposition}\label{thm_tightness_restate}
For $q \ge 1 + \binom{ n}{  k+ \mu}$, there exists a $k\times n$ matrix $G = (g_i)_{i \in N}$ over $\mathbb{F}_q$ satisfying the following conditions:
\begin{itemize}
    \item[(1)] $\sum_{i \in B_{a}^{(l)}} g_i = 0$ for  $1\leq l\leq k$ and $1\leq a\leq \delta-1$.
    \item[(2)] $\text{rank}(g_{s^{(1)}},\cdots,g_{s^{(k)}}) = k$.
    \item[(3)] For any $M \subseteq N$ with $\left| M \right| = k+\mu$, $\text{rank}(M) = k$.
\end{itemize}
\end{proposition}

\begin{IEEEproof} We first define the set of variables.
Let
\begin{equation}\{X^{(l)}, X^{(l)}_{ab}\mid 1\leq l\leq k, 1\leq a\leq \delta-1, 1\leq b\leq r-1\}\label{eq5}\end{equation}
be a set of $k((r-1)(\delta -1)+1)$ column vectors of length $k$ where components of each vector are variables over $\mathbb{F}_q$. For $1 \le l \le k$ and $1 \le a \le \delta -1$, let $$X_{s^{(l)}_{ar}} = - (X_{s^{(l)}} + X_{s^{(l)}_{a1}} + \cdots + X_{s^{(l)}_{a r-1}})\;.$$
Then at each evaluation of the variables in (\ref{eq5}),  $G(X) = (X_i)_{i \in N}$ is a $k \times n$ matrix over $\mathbb{F}_q$ satisfying the condition (1). Our goal is to find a evaluation of variables in (\ref{eq5}) at which the matrix $G(X)$ also satisfies the condition (2) and (3).

We call a set $F \subseteq N$ is free if the submatrix $G(X)|_F=(X_i)_{i\in F}$ could be any $k\times |F|$ matrix over $\mathbb{F}_q$ as the variables in (\ref{eq5}) range over   $\mathbb{F}_q$. Obviously, if $F \subseteq N$ is free and $\left| F \right| =k$, then $\det (G(X)|_F)$ is a nonzero polynomial since it has nonzero evaluations.

For any $M \subseteq N$ with $\left| M \right| = k+ \mu$, denote $M_i = M \cap N_i$ for $1\leq i\leq k$. If $M_i \ne \emptyset$, in the following we will find  $M^\prime_i \subseteq M_i$ such that
\begin{itemize}
    \item[(1)] $\left| M^\prime_i \right| \ge \left| M_i \right| - \lfloor \frac{\left| M_i \right| -1}{r} \rfloor$.
    \item[(2)] $M^\prime_i$ is a free set.
\end{itemize}
Specifically, there are two cases to be considered.

\it Case 1. \normalfont $s^{(i)} \in M_i$, then at most $\lceil \frac{\left| M_i \right| -1}{r} \rceil$  out of the $\delta-1$ sets  $B_1^{(i)}, \cdots, B_{\delta-1}^{(i)}$ are contained in $M_i$.
In this case, $M^\prime_i$ is constructed from $M_i$ by deleting one point (for example, the bottom point) from each of the $B_a^{(i)}$ which is contained in $M_i$.
Clearly $M^\prime_1$ is a free set and
\begin{equation*}
\left| M^\prime_i \right| \ge \left| M_i \right| - \lceil \frac{\left| M_i \right| -1}{r} \rceil.
\end{equation*}

\it Case 2. \normalfont $s^{(i)} \notin M_i$, then $ \left| B_a^{(i)} \cap M_i \right| \le r$ for $1\le a \le \delta -1$.
Without loss of generality, let $B_1^{(i)}, \cdots, B_{\xi}^{(i)}$ be the sets satisfying $ \left| B_a^{(i)} \cap M_i \right| = r$ for $1\leq a\leq \xi$.
Then we have $ \xi \le \lceil \frac{\left| M_i \right|}{r} \rceil$.
Construct the set $M^\prime_i$  by deleting one element from each of $B_2^{(i)}, \cdots, B_{\xi}^{(i)}$.
We can see that $M^\prime_i$ is free and
\begin{eqnarray*}
\left| M^\prime_i \right| & \ge & \left| M_i \right| - \lceil \frac{\left| M_i \right|}{r} \rceil +1\\
  & \ge & \left| M_i \right| - \lfloor \frac{\left| M_i \right| -1}{r} \rfloor.
\end{eqnarray*}

Now based on all the free sets $M'_i$, we construct
$$M^\prime = \bigcup_{\substack{1 \le i \le k \\ M_i \ne \emptyset}} M^\prime_i,$$
then $M^\prime$ is also a free set and
\begin{eqnarray*}
\left| M^\prime \right| & = & \sum_{\substack{1 \le i \le k \\ M_i \ne \emptyset}} \left| M^\prime_i \right| \\
& \ge & \sum_{\substack{1 \le i \le k \\ M_i \ne \emptyset}} (\left| M_i \right| - \lfloor \frac{\left| M_i \right| -1}{r} \rfloor)\\
& = & k+ \mu - \sum_{\substack{1 \le i \le k \\ M_i \ne \emptyset}} \lfloor \frac{\left| M_i \right| -1}{r} \rfloor.
\end{eqnarray*}
Note that
\begin{eqnarray*}
\sum_{\substack{1 \le i \le k \\ M_i \ne \emptyset}} \lfloor \frac{\left| M_i \right| -1}{r} \rfloor & \le & \frac{1}{r} \sum_{\substack{1 \le i \le k \\ M_i \ne \emptyset}} (\left| M_i \right| -1) \\
& = & \frac{1}{r} (k + \mu - \sum_{\substack{1 \le i \le k \\ M_i \ne \emptyset}} 1)\\
& \le & \frac{1}{r} (k + \mu - \lceil \frac{k+\mu}{r(\delta-1)+1} \rceil),
\end{eqnarray*}
where the last inequality holds because there are at least $\lceil \frac{k+\mu}{r(\delta-1)+1} \rceil$ sets out of $M_1, \cdots, M_k$ are nonempty.
Then it has $\sum_{\substack{1 \le i \le k \\ M_i \ne \emptyset}} \lfloor \frac{\left| M_i \right| -1}{r} \rfloor < \mu +1$ from Lemma \ref{lem_for_proof_of_tightness} below.

Since $\sum_{\substack{1 \le i \le k \\ M_i \ne \emptyset}} \lfloor \frac{\left| M_i \right| -1}{r} \rfloor$ is an integer, thus
 $\left| M^\prime \right| \ge k + \mu - \mu = k$.
It follows that for any $M \subseteq N$ with $\left| M \right| = k+ \mu$, one can find $S_M \subseteq M^\prime \subseteq M$ such that $S_M$ is free and $\left| S_M \right| = k$.

Let
$$ f(X) = \det (X_{s^{(1)}},\cdots,X_{s^{(k)}}) \prod_{\substack{M \subseteq N \\ \left| M \right| = k + \mu}} \det(G(X)|_{S_M}).$$
Then $f(X)$ is a nonzero polynomial and the degree of each variable is at most $\binom{n}{k + \mu} +1$.
Therefore by Schwartz-Zippel Lemma, $f(X)$ is nonzero at some evaluation of the variables, and this evaluation in turn gives the desired matrix $G$.

\end{IEEEproof}

\begin{lemma}\label{lem_for_proof_of_tightness}
$$\frac{1}{r}(k+\mu - \lceil \frac{k+\mu}{r(\delta-1)+1}\rceil) < \mu+1,$$
where $ \mu = \lceil \frac{(k-1)(\delta-1)+1}{(r-1)(\delta-1)+1} \rceil -1 .$
\end{lemma}
\begin{IEEEproof}
It is equivalent to prove that $k < (r-1)\mu +r + \lceil \frac{k+\mu}{r(\delta-1)+1} \rceil$.
Note that
\begin{eqnarray*}
\mu & \ge & \frac{(k-1)(\delta-1)+1}{(r-1)(\delta-1)+1} -1 \\
& =  & \frac{(k-r)(\delta-1)}{(r-1)(\delta-1)+1}.
\end{eqnarray*}
Then
\begin{eqnarray*}
& & (r -1 ) \mu + r +\lceil \frac{k+\mu}{r(\delta-1)+1} \rceil\\
& \ge & (r-1) \cdot \frac{(k-r)(\delta-1)}{(r-1)(\delta-1)+1} +r + \frac{k + \frac{(k-r)(\delta-1)}{(r-1)(\delta-1)+1}}{ r(\delta-1)+1} \\
& = & k + \frac{r}{r(\delta-1)+1}\\
& > & k.
\end{eqnarray*}

\end{IEEEproof}

\begin{proposition}\label{prop_distance_ex3}
When $q > \binom{n}{k+\mu_k}$, there exists a generator matrix $G(X)$ over $\mathbb{F}_q$ such that the minimum distance of $\mathcal{C}$ satisfies $d \ge n-k+1-\mu_k$.
\end{proposition}
\begin{IEEEproof}
From Lemma \ref{rank(n-d+1)=k}, we finish the proof by showing any submatrix of $G(X)$ containing $k+\mu_k$ columns has rank $k$.
Let $T$ be a set of $k+ \mu_k$ cross points in the grid of Fig. \ref{fig_square_code}.
Suppose $T$ contains $\rho$ entire columns each of which consists of $r+1$  points.
Besides, suppose $T$ contains at lest $\sigma$  points in each of the remaining columns.
Note that either of $\rho$ and $\sigma$  could be zero.

\begin{figure}[htpb]
\centering

\begin{tikzpicture}

\fill [blue!70] (0,0) rectangle (1,3);
\fill [blue!70] (1,0) -- (1,1.5) .. controls (1.5,1.8) and (2,1) .. (3,1.5) -- (3,0) -- (1,0);
\draw (0,0) rectangle (3,3);
\draw (0,0) rectangle (1,3);
\draw (1,1.5) .. controls (1.5,1.8) and (2,1) .. (3,1.5);
\draw [->] (3.3,2) -- (2,.8);

\node at (.5,3.3) {\small $\rho$ columns};
\node at (3.5,2.2) {\small at leat $\sigma$ points in each column};

\node at (-2,0) {};

\end{tikzpicture}

\caption{$T$ is represented by the shadow part.}
\label{fig_T}

\end{figure}
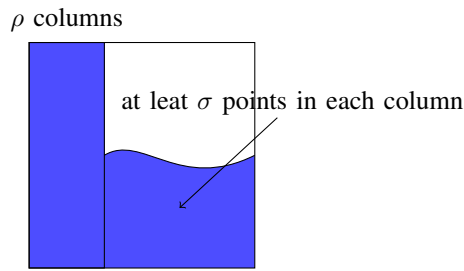

Fig. \ref{fig_T} gives a simple instance of $T$ by assuming the points of $T$ in each column are consecutive.
Actually, intervals are allowed.
Then we have $\left| T \right| \ge \rho(r+1) + \sigma (r+1 -\rho) = (\rho+\sigma)(r+1) - \rho \sigma$.

Similar to the proof of Theorem \ref{thm_tightness}, taking the vectors in $X$ as variables, we define a subset $T^\prime \subseteq T$ as a free set if each vector in $\{x_t\}_{t\in T^\prime}$ can independently range over $\mathbb{F}_q^{k}$ as the variables vary.
Here $x_t$ denotes the vector associated with the point $t$.
Next we will find a free set in $T$ containing at least $k$ points.
As in Fig. \ref{fig_tilde_T}, we obtain $T^\prime \subseteq T$ by deleting one point from each of the entire columns and additionally the column containing $\sigma $ points.

\begin{figure}[htbp]
\centering

\begin{tikzpicture}

\fill [blue!70] (0,0) rectangle (1,2.8);
\fill [blue!70] (1,0) -- (1,1.5) .. controls (1.5,1.8) and (2,1) .. (3,1.5) -- (3,0) -- (1,0);
\fill [white] (2.2,0) -- (2.2,3) -- (2.4,3) -- (2.4,0) -- (2.2,0);
\draw (0,0) rectangle (3,3);
\draw (0,0) rectangle (1,3);
\draw (1,1.5) .. controls (1.5,1.8) and (2,1) .. (3,1.5);
\draw [->] (3.3,2) -- (2.3,.8);
\draw [->] (.5,3.2) -- (0.4,2.9);

\node at (1.5,3.3) {\small delete one point from each column};
\node at (3.5,2.2) {\small delete the $\sigma$-column};

\end{tikzpicture}

\caption{$T^\prime$ is represented by the shadow part.}
\label{fig_tilde_T}
\end{figure}

It is evident that $T^\prime$ is a free set.
The size of $T^\prime$ is $\left| T \right| - (\rho+\sigma)$.
Furthermore, it must have $\rho+\sigma \le \mu_k$.
Otherwise, if $\rho+\sigma > \mu_k$, then
\begin{eqnarray*}
f(\rho+\sigma) &  \ge & \rho+\sigma+k \\
& > & \mu_k + k\\
& = & \left| T \right| \\
& \ge & (\rho+\sigma)(r+1)-\rho \sigma,
\end{eqnarray*}
where the first inequality follows from the definition of $\mu_k$.
Thus
\begin{eqnarray*}
0 & < & f(\rho+\sigma) - (\rho+\sigma)(r+1) +\rho \sigma \\
& = & \begin{cases} \rho \sigma - \frac{(\rho + \sigma)^2}{4}, \text{ if } 2 \mid \rho + \sigma \\ \rho \sigma - \frac{(\rho + \sigma)^2-1}{4}, \text{ if } 2 \nmid \rho + \sigma \end{cases},
\end{eqnarray*}
which is impossible.

It follows $| T^\prime | \ge k$. Then for any $T$ with $\left| T \right| = k+ \mu_k$, we can find  $S_T \subseteq T^\prime\subseteq T$ such that $S_T$ is a free set and $\left| S_T \right| = k$.

Let $$ f(X) = \prod_{\substack{T \subseteq [n] \\ \left| T \right| = k+\mu_k}} \det (G(X)|_{S_T}).$$
Existence of the free set $S_T$ indicates that $\det(G(X)|_{S_T})$ is a nonzero polynomial.
Then $f(X)$ is a nonzero polynomial and the degree of each variable is at most $\binom{n}{k+\mu_k}$.
By Schwartz-Zippel Lemma, $f(X)$ is nonzero at some evaluation of the variables when $q > \binom{n}{k+\mu_k}$, and this evaluation gives the generator matrix $G(X)$ from which the linear code has minimum distance at least $n-k+1-\mu_k$.
\end{IEEEproof}

\end{document}